# Statistics of Quantum Turbulence in Superfluid He

**V.S. L'vov · A. Pomyalov**

**Abstract** Based on our current understanding of statistics of quantum turbulence as well as on results of intensive ongoing analytical, numerical and experimental studies we overview here the following problems in the large-scale, space-homogeneous, steady-state turbulence of superfluid $^4$He and $^3$He: 1) Energy spectra of normal and superfluid velocity components; 2) Cross-correlation function of normal and superfluid velocities; 3) Energy dissipation by mutual friction and viscosity; 4) Energy exchange between normal and superfluid components; 5) High-order statistics and intermittency effects. The statistical properties are discussed for turbulence in different types of flows: coflow of $^4$He; turbulent $^3$He with the laminar normal fluid; pure superflow and counterflow in $^4$He.

**Keywords** Energy spectra · Intermittency · Velocity decoupling

## 1 Introduction

Turbulence in superfluid helium was predicted first by Feynman in 1955 [1] and found experimentally (in $^4$He counterflow) by Vinen in 1957 [2]. This phenomenon is unusual and presents a challenge to fluid dynamicists because it consists of two coupled, interpenetrating turbulent fluids: the first is inviscid with quantized vorticity, the second is viscous with continuous vorticity [3, 4].

In superfluid He, vorticity is constrained to vortex line singularities of fixed circulation $\kappa = h/M$, where $h$ is Plancks constant, and $M$ is the mass of the relevant boson ($M$ is equal to the mass of $^4$He atom or two masses of $^3$He atom, creating a Cooper pair). Quantization of vortex lines creates two additional scales: the vortex core radius $a_0$ (about $10^{-8}$ cm in $^4$He and $8 \cdot 10^{-6}$ cm in $^3$He) and the mean intervortex distance $\ell = 1/\sqrt{\mathcal{L}}$, where $\mathcal{L}$ is the vortex-line

V.S. L'vov · A. Pomyalov
Department of Chemical Physics, Weizmann Institute of Science, 76100 Rehovot, Israel
E-mail: victor.lvov@gmail.com

density (VLD). In typical experiments $\mathcal{L}$ is about $(10^4 \div 10^6)\,\mathrm{cm}^{-2}$, giving $\ell \sim (10^{-3} - 10^{-2})\,\mathrm{cm}$. This is much larger than the core radius $a_0$, but much smaller than the outer scale of turbulence $\Delta \sim (1-100)$cm. In this review we consider semi-classical region of scales $\ell \lesssim R \lesssim \Delta$.

It is commonly believed that in mechanically driven turbulence of superfluids the mean velocities of normal and superfluid components coincide, $\boldsymbol{V}_\mathrm{n} = \boldsymbol{V}_\mathrm{s}$ (so-called coflow), perhaps except for a narrow region near walls, and that mutual friction between the normal and superfluid components couples also their turbulent fluctuations almost at all scales $\ell \lesssim R \lesssim \Delta$ [4–7]. The similarity and difference between statistics of turbulence in classical fluid and in superfluids is discussed in Sec. 2. First, in Sec. 2.1 we recall the Richardson-Kolmogorov picture of classical turbulence and intermittency effects. The manifestation of the intermittency in quantum turbulence [7,8] is discussed in following Sec. 2.2, while in Sec. 2.3 we consider turbulent kinetic energy balance in normal and superfluid components.

Next Sec. 3 is devoted to clarification of turbulent energy spectra of the superfluid component in the limiting case, when normal fluid component can be considered at rest [9,10]. This limit is inspired by properties of $^3$He, which has very high kinematic viscosity of the normal fluid component. We show that in this case energy dissipation due to the mutual friction at all scales plays crucial role, leading to dramatic suppression of the energy spectra compared to the classical and mechanically driven quantum turbulence.

The situation in thermally driven turbulence, Sec. 4, is much more complicated than in two previous limiting cases. We demonstrate that non-vanishing counterflow velocity leads to scale-dependent partial decorrelation turbulent velocity fluctuations of two components. As a result, the relative role of mutual friction and molecular viscosity depends on the energy balance at a given scale $r$ (or wave number $k \simeq \pi/r$) and the resulting energy spectrum share similarities with the spectra in both the classical and quantum $^3$He turbulence.

The background on experiments in the counterflow turbulence, based on Refs. [11,12], is given in Sec. 4.1. Simple analytical model [11] of turbulent energy spectra in the presence of counterflow is presented in Sec. 4.2. The resulting second-order velocity structure functions are compared with experiment [13,14] in Sec. 4.3.

In our overview we are not pretending to present a comprehensive description of all results in the problem of statistical description of quantum turbulence. They can be found in Refs. [3–7] and numerous (not cited for shortness) publications. Our goal is modest: we present here our current understanding of this problem, attempting to sketch it as a whole, without going into too many details, and aiming to make it understandable not only to experts in turbulence, but to a wider community in quantum fluids and solids.

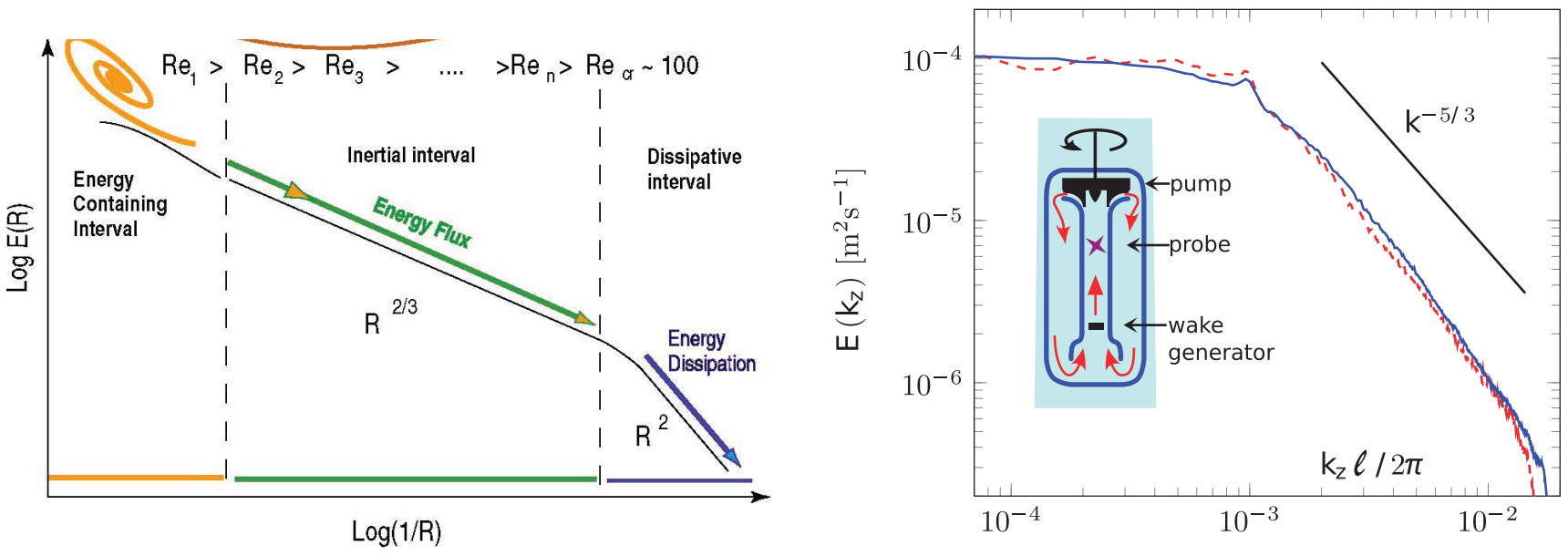

**Fig. 1** Panel a: Sketch of the Richardson-Kolmogorov energy cascade. Panel b: Energy spectrum measured in the TOUPIE wind tunnel (shown in Inset) below the superfluid transition (solid blue line, 1.56 K< $T_\lambda$) and above $T_\lambda$ (dashed red line)[16].(Color Online)

## 2 Mechanically driven turbulence of superfluid $^4$He

### 2.1 Classical Richardson-Kolmogorov cascade and intermittency effects

Recall [15] that incompressible viscous flows of classical fluids, such as water and air, are described by the Navier-Stokes (NS) equation

$$[\partial \boldsymbol{u}/\partial t + (\boldsymbol{u}\cdot\boldsymbol{\nabla})u] - \boldsymbol{\nabla} p/\rho = \nu \nabla^2 \boldsymbol{u}\,, \quad \mathrm{Re} = V_{\rm T}\Delta/\nu\,, \tag{1}$$

for the velocity field $\boldsymbol{u}(\boldsymbol{r},t)$. Here $p$ is the pressure, $\rho$ is density of the fluid, $\nu$ is its kinematic viscosity and $V_{\rm T}$ is the root-mean-square turbulent velocity fluctuations. When Reynolds number Re is very large, the nonlinear term in Eq. (1) dominates over the viscous term and largest $\Delta$-scale eddies are unstable and give birth to smaller scale eddies, which, being unstable too, generate further smaller eddies, and so on. This Richardson-Kolmogorov cascade, sketched in Fig. 1a, transfers energy toward viscous scale $\eta$, at which the nonlinear and viscous forces in Eq. (1) approximately balance each other.

Kolmogorov in 1941 (K41) assumed [15] that small scale $R \ll \Delta$ turbulence may be approximately considered as homogeneous and isotropic, with universal statistics that depends in the inertial interval of scales $\Delta > R > \eta$ only on one relevant parameter – mean energy flux $\varepsilon$ (with the dimensionality $[\varepsilon]$ =cm$^2$/s$^3$). If so, the velocity of $R$-scale eddies may be estimated as $(\varepsilon R)^{1/3}$ and velocity structure functions $S_n(R)$, which are dominated in the inertial interval by $R$-eddies, scale as follows:

$$S_n(R,t) = \langle |\boldsymbol{v}(\rho + \boldsymbol{R}, t) - \boldsymbol{v}(\boldsymbol{r},t)|^n\rangle \simeq (\varepsilon R)^{n/3}\ . \tag{2a}$$

Another way to characterize the energy distribution in the wavenumber $k$-space ($k \simeq \pi/R$) is one-dimensional energy spectrum $E(k,t)$, normalized such that energy density per unite mass is defined as $E(t) = \frac{1}{2}\int |\boldsymbol{u}|^2 d\boldsymbol{r}/\int d\boldsymbol{r} = \int_0^\infty E(k,t)dk$. K41 dimensional reasoning gives then the celebrated 5/3-law:

$$E(k) \simeq \varepsilon^{2/3} k^{-5/3}\ . \tag{2b}$$

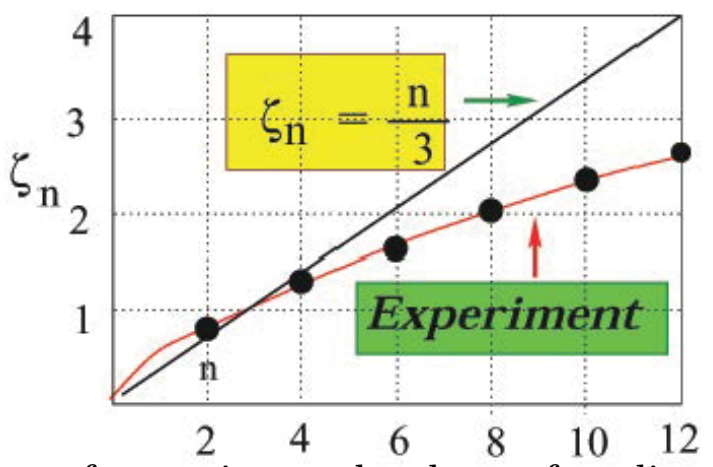

| n | n/3 | $\zeta_n$ |
|---|---|---|
| 2 | 0.67 | 0.70 |
| 4 | 1.33 | 1.20 |
| 6 | 2.00 | 1.62 |
| 8 | 2.67 | 2.00 |
| 10 | 3.33 | 2.36 |
| 12 | 4.00 | 2.68 |

**Fig. 2** Comparison of experimental values of scaling exponents [15] $\zeta_n$ (green) and their K41 estimates (yellow). (Color online).

This simple, one-line derivation agrees surprisingly well with numerous numerical simulations of Eq. (1), observations in the Nature and laboratory experiments, including liquid $^4$He below and above the $\lambda$-point $T_\lambda \simeq 2.17\,$K (transition to superfluidity), see Refs. [9, 10] and Fig. 1b.

However, the situation is not so simple. Accurate experiments and numerical simulations show that scaling exponents $\zeta_n$ of $S_n(R) \propto R^{\zeta_n}$ definitely deviate from K41 values (2a), $\zeta_n^{\text{K41}} = n/3$, see Fig. 2 .

In attempt to amend this discrepancy, Kolmogorov in 1962 [15] used the analogy of random breaking of stones with that of eddies, and assumed Gaussian statistics of $\ln \varepsilon(R)$. In the resulting phenomenological *K62 log-normal model* $\zeta_n = \dfrac{n}{3} - \dfrac{\mu_2}{18} n(n-3)$, which with $\mu_2 \simeq 0.35$. Unfortunately, K62 and all other phenomenological models of multi-scaling (intermittency) are fully disconnected from the Navier-Stokes dynamics (1).

On the other hand, in dynamical models of intermittency, like shell models [17, 18], $\boldsymbol{u}(\boldsymbol{k}, t)$ is replaced in the entire shell of wavenumbers $k_m = 2^m k_0 < k < k_{m+1}$ by only one complex *shell velocity* $u_m(t)$ and mimics the nonlinear term in NS Eq. (1) in the wavevector $\boldsymbol{k}$-representation. For instance, in Sabra shell model [18] $\text{NL}_m\{\boldsymbol{u}_{m'}\}$ is represented as follows:

$$\Big(\frac{d}{dt} + \nu k_m^2\Big) u_m + \text{NL}_m\{\boldsymbol{u}_{m'}\} = 0\,, \tag{3a}$$

$$\text{NL}_m\{\boldsymbol{u}_{m'}\} = i k_m \Big(2a\, u_{m+2} u^*_{m+1} + b\, u_{m+1} u^*_{m-1} - \frac{c}{2}\, u_{m-1} u_{m-2}\Big)\ . \tag{3b}$$

For $a + b + c = 0$ it conserves the total kinetic energy $E = \sum_m |u_m|^2$ in the inviscid limit, $\nu = 0$ and for $b = -0.5\,a$ [17, 18] the scaling exponents are in good agreement with experimentally known values of $\zeta_n$ and well describe other statistical details of classical hydrodynamic turbulence.

## 2.2 Intermittency enhancement in superfluid turbulence

Large scale turbulence in superfluid $^4$He can be described by the Landau-Tirsza two-fluid model with interpenetrating normal and superfluid components, each with its own density $\rho_\text{n}$, $\rho_\text{s}$ and its own velocity fields $\boldsymbol{u}_\text{n}(\boldsymbol{r}, t)$, $\boldsymbol{u}_\text{s}(\boldsymbol{r}, t)$. The coarse-grained Hall-Vinen [19]-Bekarevich-Khalatnikov[20] (HVBK) equations, that describe this system, have a form of two NS Eqs. (1) for $\boldsymbol{u}_{\rm n}(\boldsymbol{r},t)$ and $\boldsymbol{u}_{\rm s}(\boldsymbol{r},t)$:

$$\rho_{\rm n}[\partial \boldsymbol{u}_{\rm n}/\partial t + (\boldsymbol{u}_{\rm n}\cdot\boldsymbol{\nabla})u_{\rm n}] - \boldsymbol{\nabla} p_{\rm n} = \rho_{\rm n}\nu_{\rm n}\nabla^2\boldsymbol{u}_{\rm n} - \boldsymbol{F}\,, \tag{4a}$$

$$\rho_{\rm s}[\partial \boldsymbol{u}_{\rm s}/\partial t + (\boldsymbol{u}_{\rm s}\cdot\boldsymbol{\nabla})u_{\rm s}] - \boldsymbol{\nabla} p_{\rm s} = \rho_{\rm s}\nu_{\rm s}\nabla^2\boldsymbol{u}_{\rm s} + \boldsymbol{F}\,, \tag{4b}$$

coupled by the mutual friction force $\boldsymbol{F}$ between two components of He, which may be approximated as [9]:

$$\boldsymbol{F} \simeq \alpha(T)\rho_{\rm s}(\boldsymbol{u}_{\rm s} - \boldsymbol{u}_{\rm n})\Omega\,, \quad \Omega = \kappa\mathcal{L}\,. \tag{4c}$$

Here $\alpha(T)$ is the dimensionless mutual friction coefficient. The dissipative term with the Vinen's effective superfluid viscosity $\nu_{\rm s} \sim \alpha\kappa$ was added in [8] to account for the energy dissipation on the intervortex scale $\ell$ due to vortex reconnections and similar effects.

In numerical studies of the superfluid turbulence in Refs. [8, 10, 21], we replaced HVBK Eqs. (4) by their Sabra-shell model counterparts Eqs. (3)

$$\Big(\frac{d}{dt} + \nu k_m^2\Big)u_m^{\rm n} + {\rm NL}_m\{\boldsymbol{u}_{m'}^{\rm n}\} = -\frac{F_{\rm m}}{\rho_n}\,, \quad \Big(\frac{d}{dt} + \nu k_m^2\Big)u_m^{\rm s} + {\rm NL}_m\{\boldsymbol{u}_{m'}^{\rm s}\} = \frac{F_{\rm m}}{\rho_s}\,, \tag{5}$$

coupled by the mutual friction force $F_{\rm m} = \alpha(T)(u_m^{\rm s} - u_m^{\rm n})\Omega$, in which instantaneous $\Omega$ is estimated as: $\Omega = \sum_m k_m^2|u_m^{\rm s}|^2$.

Comprehensive numerical simulations of Eqs. (5) with *eight decades* of the inertial interval [8, 10, 21] allowed detailed comparison of classical and superfluid turbulent statistics. In particular, in Ref. [8] we demonstrated that scaling exponents $\zeta_n$ of $S_n(R) \propto R^{\zeta_n}$ in superfluid $^4$He turbulence are the same as in classical turbulence for temperatures close to the superfluid transition $T_\lambda$ and also for $T \ll T_\lambda$. This result may be easily rationalized: at very low temperatures the contribution of normal fluid component is negligible, while for $T$ close to $T_\lambda$, it gives the dominant contribution. In both cases, the equation for the dominant component practically coincides with NS Eq. (1). This observation agrees with earlier numerical study of intermittent exponents [22] and with low-temperature experiments [22,23] for $T \lesssim 0.7\,T_\lambda$, when $\rho_s/\rho_n > 5.7$.

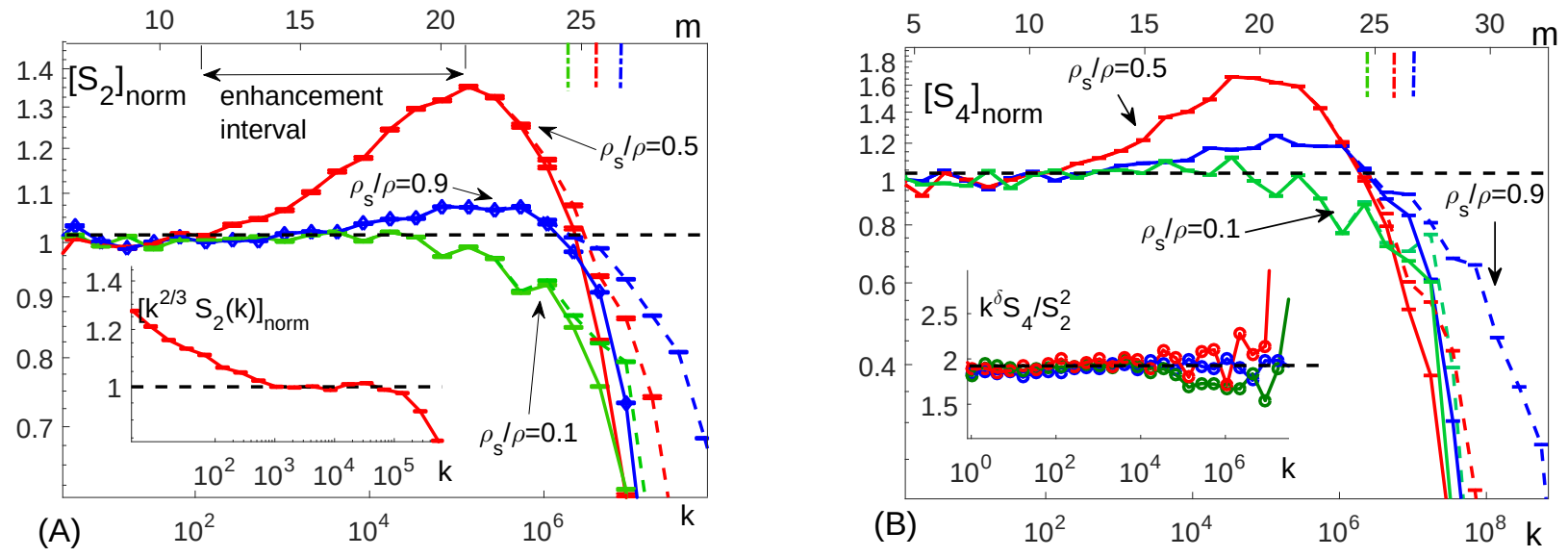

**Fig. 3** Panel A: The compensated structure functions $S_2^{\rm n,s}(k_m)$ for $\rho_{\rm s}/\rho = 0.1, 0.5$ and 0.9, normalized by their corresponding mean values over shells with $m = 7-14$. Inset: $k^{2/3}\,S_2(k_m)$ normalized by its mean value over $m = 15-20$ for $\rho_{\rm s}/\rho = 0.5$. Solid lines – normal component, dashed lines – superfluid components. The upper axis is marked with the shell numbers, the lower axis shows the corresponding wavenumbers. Panel B: The normalized compensated $S_4^{\rm n,s}(k_m)$. Inset: Collapse of flatness, compensated using numerical exponents $\delta = 2\xi_2^{\rm num} - \xi_4^{\rm num} = 0.184$. The horizontal black dashed lines serve to guide the eye only. Color code is the same in both panels. Vertical dot-dashed lines denote the intervortex scale $\ell$.[8] (Color Online).

The situation is less trivial in the temperature range $0.8T_\lambda < T < 0.9T_\lambda$, for which $\rho_s \approx \rho_n$ and contribution of mutual friction becomes important. In this case, there exists a wide (up to three decades) range of scales, adjacent to the viscous cutoff, in which scaling of $S_n(R)$ is shallower than in the classical turbulence. In particular, the apparent scaling of the energy spectrum is close to K-41 prediction, see red lines in Fig. 3A. However, this does not indicate suppression of intermittency. On the contrary, the scaling exponents $\zeta_n$ with $n > 3$ in this range of scales deviatate further from the K41 prediction, as confirmed by scaling of flatness, see Fig. 3B. While interpretation of these observations as a kind of a bottleneck effect in the two-fluid system [8] looks plausible, this phenomenon definitely calls for further investigation by direct numerical simulations of the HVBK Eqs. (4) and in laboratory experiments. Both ways are nowadays realistic.

### 2.3 Energy spectra, velocity cross-correlation and energy balance

The energy spectra, obtained with the Sabra-shell model form (5) of HVBK Eqs. (4), for temperatures $T \gtrsim T_\lambda/2 \simeq 1.08\,$K, demonstrate in a wide inertial interval of scales the same intermittent behavior $\mathcal{E}(k) \propto k^{-1.72}$ as in classical hydrodynamic turbulence [21]. The high $k$-region of such spectra, adjacent to the viscous cutoff, are shown in Fig. 4a for $T = 1.07, 1.3, 1.8$ K and $T = 2.16\,$K. At lowest $T = 1.07\,$K (blue lines) the normal fluid spectrum terminates about two decades earlier than its superfluid counterpart. This dif-ference decreases for $T = 1.3\,\mathrm{K}$(cyan lines) and $T = 1.8\,\mathrm{K}$ (green lines), while for $T = 2.16\,K$(close to $T_\lambda$), superfluid spectrum terminates earlier than the normal fluid one. The main reason for such a behavior is the relation between the kinematic viscosity $\nu_\mathrm{n}$ of normal fluid and the effective Vinen's viscosity $\nu'_\mathrm{s}$ of the superfluid. At very low temperatures $\nu_\mathrm{n} \gg \nu'_\mathrm{s}$, see Fig. 4b, at moderate $T = 1.8\,\mathrm{K}$ the viscosities are close, while for $T = 2.16\,K$ $\nu'_\mathrm{s} > \nu_\mathrm{n}$.

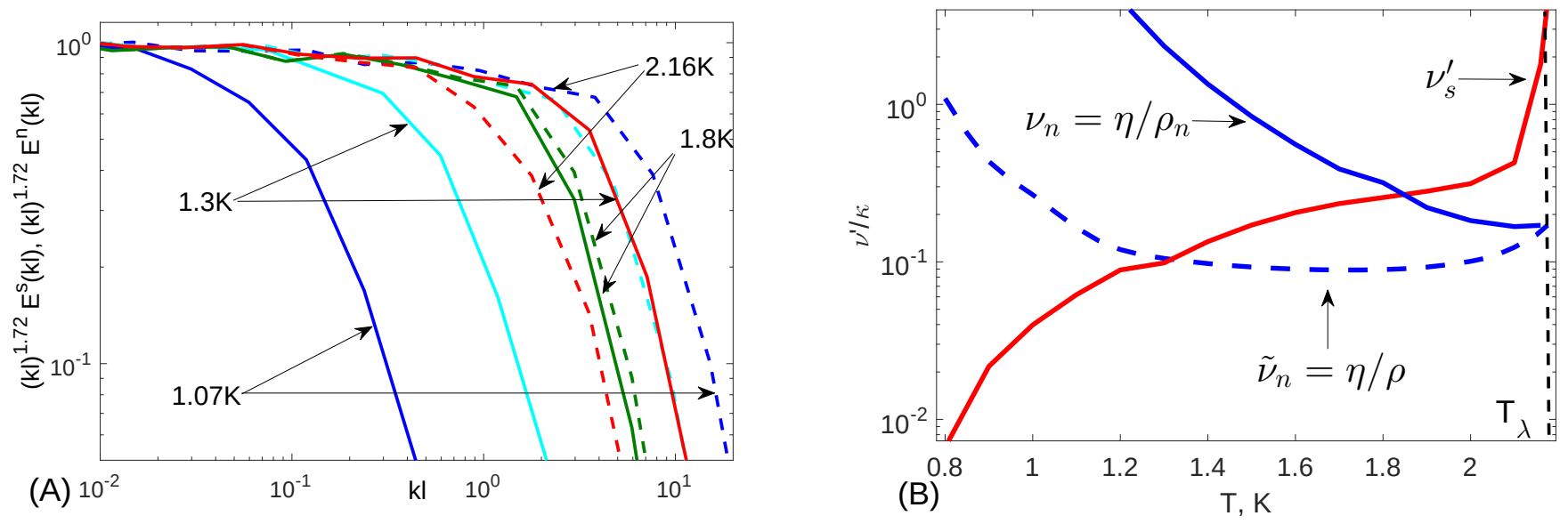

**Fig. 4** Panel a: Energy spectra $\mathcal{E}_n(k)$ and $\mathcal{E}_s(k)$ of the normal fluid(solid lines) and superfluid (dashed lines), compensated by the anomalous scaling behavior $k^{1.72}$, for different $T$. Panel b: Temperature dependence of the kinematic viscosity $\nu_n$ of the normal (blue line) and effective viscosity $\nu'_s$ of superfluid (red line) components. Dashed line – normal fluid kinematic viscosity, normalized by the total $^4$He density.[21](Color Online)

More detailed analysis of the energy spectra requires knowledge about correlation between normal and superfluid velocities. It is often assumed (see e.g. Ref. [4]) that the normal and superfluid velocities are "locked" in the sense that $\boldsymbol{u}_\mathrm{n}(\boldsymbol{r}, t) = \boldsymbol{u}_\mathrm{s}(\boldsymbol{r}, t)$. For quantitative understanding to which extent this assumption is statistically valid, we introduce 1D cross-velocity correlation function $\mathcal{E}_\mathrm{ns} \propto \langle \boldsymbol{u}_\mathrm{n}(\boldsymbol{k}) \cdot \boldsymbol{u}_\mathrm{s}(\boldsymbol{k})\rangle$, normalized such that $\int \mathcal{E}_\mathrm{ns}(k)\, dk = \langle \boldsymbol{u}_\mathrm{n}(\boldsymbol{r}, t)\boldsymbol{u}_\mathrm{s}(\boldsymbol{r}, t)\rangle /2$. If, for example, motions of the normal and the superfluid components at a given $k$ are completely correlated, then $\mathcal{E}_\mathrm{ns}(k) = \mathcal{E}_\mathrm{n}(k) = \mathcal{E}_\mathrm{s}(k)$. If this is true for all scales, then $\boldsymbol{u}_\mathrm{n}(\boldsymbol{r}, t) = \boldsymbol{u}_\mathrm{s}(\boldsymbol{r}, t)$.

It is natural to normalize $\mathcal{E}_\mathrm{ns}$ by the normal and the superfluid energy densities, $\mathcal{E}_\mathrm{n}$ and $\mathcal{E}_\mathrm{s}$. This can be reasonably done in one of two ways:

$$\mathcal{K}_1(k) = \frac{2\,\mathcal{E}_\mathrm{ns}(k)}{\mathcal{E}_\mathrm{n}(k) + \mathcal{E}_\mathrm{s}(k)}\,, \quad \text{or} \quad \mathcal{K}_2(k) \equiv \frac{\mathcal{E}_\mathrm{ns}(k)}{\sqrt{\mathcal{E}_\mathrm{n}(k) \cdot \mathcal{E}_\mathrm{s}(k)}}\,. \tag{6}$$

Both cross-correlations are equal to unity for fully locked superfluid and normal velocities, and both vanish if the velocities are statistically independent. However, if $\boldsymbol{u}_\mathrm{n}(\boldsymbol{r}, t) = C\boldsymbol{u}_\mathrm{s}(\boldsymbol{r}, t)$, with $C \neq 1$, then $\mathcal{K}_1(k) = 2C/(C^2 + 1) < 1$, but still $\mathcal{K}_2(k) = 1$. In any case $\mathcal{K}_1(k) \leqslant \mathcal{K}_2(k)$.

Cross-correlations $\mathcal{K}_1(k)$ and $\mathcal{K}_2(k)$ [21] are shown in Fig. 5 for different $T$. At first glance, it is surprising that $\mathcal{K}_2(k)$ (dashed lines) $T \lesseqgtr 1.8\,\mathrm{K}$ persist for much larger wave numbers than $\mathcal{K}_1(k\ell)$. For instance, for low $T$ $v^\mathrm{n}_m \ll v^\mathrm{s}_m$ for ($1 \lesssim k_m\ell \lesssim 100$), but $v^\mathrm{n}_m(t) \propto v^\mathrm{s}_m(t)$, meaning that strongly damped normal velocity does not have its own dynamics and should be considered as "enslaved" by the superfluid velocity. The damped velocity (normal or superfluid) at any temperature $T \lesseqgtr 1.8\,\mathrm{K}$ would follow this "enslaved" dynamics.

Notice that numerical results shown in Fig. 5a agree with the analytical expression of the cross-correlation $\mathcal{E}_\mathrm{sn}(k)$ [27], which in current notations reads:

$$\mathcal{E}_\mathrm{sn}(k) = \frac{\alpha\,\Omega[\rho_\mathrm{n}\,\mathcal{E}_\mathrm{n}(k) + \rho_\mathrm{s}\,\mathcal{E}_\mathrm{s}(k)]}{\alpha\,\Omega\rho + \rho_\mathrm{n}[(\nu'_\mathrm{s} + \nu_\mathrm{n})\,k^2 + \gamma_\mathrm{n}(k) + \gamma_\mathrm{s}(k)]}\,, \ \gamma_\mathrm{s,n}(k) = \sqrt{k^3\mathcal{E}_\mathrm{s,n}(k)}. \tag{7}$$

Here $\gamma_\mathrm{s}(k)$ and $\gamma_\mathrm{n}(k)$, are the turnover frequencies of eddies in the superfluid and normal fluid components, respectively.

Strong coupling of the normal and the superfluid velocities suppresses energy dissipation and energy exchange between the components, caused by the mutual friction, proportional to $\boldsymbol{u}_\mathrm{s} - \boldsymbol{u}_\mathrm{n}$. Nevertheless, some dissipation is still there. To demonstrate this, we consider the ratio of the total injected energy to the total energy, dissipated due to the viscosity, in the normal and the superfluid components $R_\mathrm{s+n} = \dfrac{\rho_\mathrm{s}\varepsilon_\mathrm{s} + \rho_\mathrm{n}\varepsilon_\mathrm{n}}{\rho_\mathrm{s}\nu'_\mathrm{s}\,\langle|\omega_\mathrm{s}|^2\rangle + \rho_\mathrm{n}\nu_\mathrm{n}\,\langle|\omega_\mathrm{n}|^2\rangle}$, a quantity plotted in

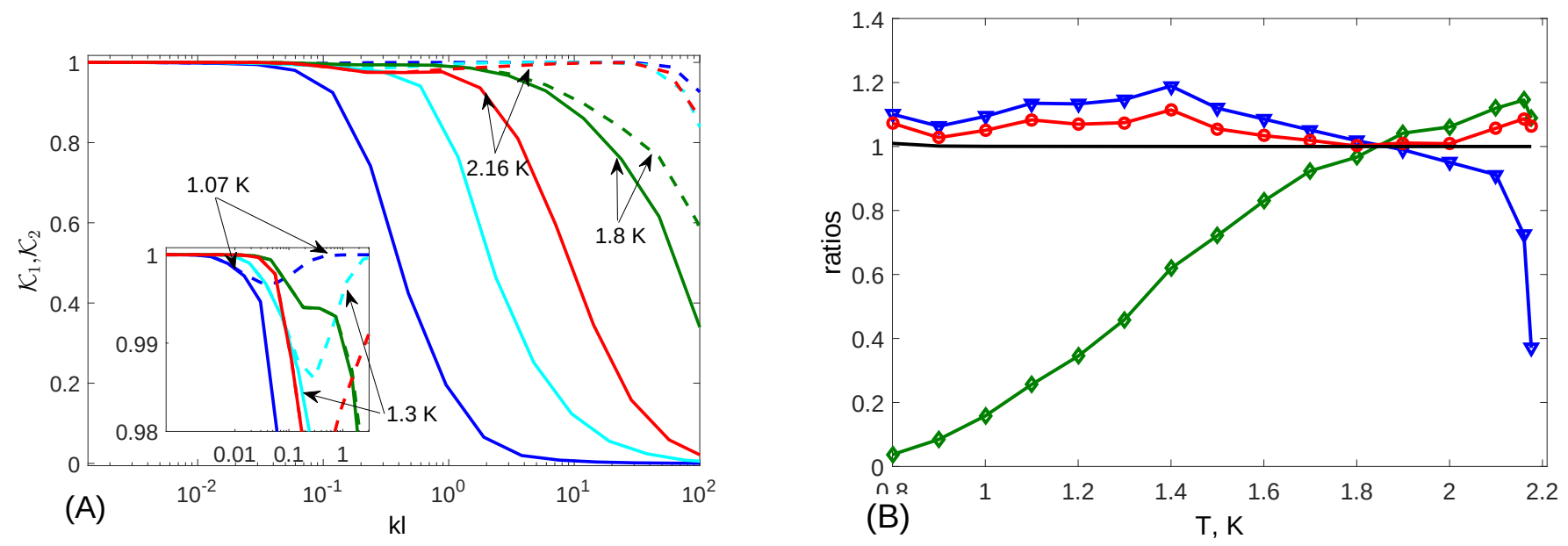

**Fig. 5** Panel a: Cross-correlation coefficients $\mathcal{K}_1(k\ell)$(solid lines), and $\mathcal{K}_2(k\ell)$, Eq. (6) (dashed lines) for different temperatures. Color code is the same as in Fig. 4a. Panel b: Temperature dependence of the ratios: $\varepsilon_{\rm s}/\varepsilon_{\rm n}$ – horizontal black line; $R_{\rm s}$ – blue line with triangles, $R_{\rm n}$ – green line with diamonds, $R_{\rm s+n}$ – red line with circles.[21] (Color Online)

Fig. 5b by red line with circles. Here $\varepsilon_{\rm n}$ and $\varepsilon_{\rm s}$ are the inertial range normal and superfluid energy fluxes. This ratio exceeds unity by about 10%, meaning that $\sim$ 10% of the injected energy is dissipated by the mutual friction. As expected, this effect disappears at $T \approx 1.8\,$K, when the effective superfluid and normal fluid kinematic viscosities are matching (and therefore $\boldsymbol{u}_{\rm s} \approx \boldsymbol{u}_{\rm n}$).

The mutual friction has significantly more important influence on the energy exchange between the components. The energy exchange can be quantified by a similar ratio defined for each fluid component, $R_{\rm s} = \dfrac{\varepsilon_{\rm s}}{\nu'_{\rm s}\left\langle |\omega_{\rm s}|^2\right\rangle}$, $R_{\rm n} = \dfrac{\varepsilon_{\rm n}}{\nu_{\rm n}\left\langle |\omega_{\rm n}|^2\right\rangle}$, shown in Fig. 5b by a green line with diamonds and by a blue line with triangles respectively. At the lowest shown temperature $T = 0.8\,$K, we have $R_{\rm n} < 0.1$ meaning that only about 10% of the energy density (per unit mass) which is dissipated by the normal fluid component comes from the direct energy input. The rest $\simeq$ 90% of the energy density, dissipated by viscosity (at large $k$), was transferred from the superfluid component by the mutual friction. As expected, there is no energy exchange between the components at $T \approx 1.8\,$K, when $\nu_{\rm n} = \nu'_{\rm s}$ and $\boldsymbol{u}_{\rm s} = \boldsymbol{u}_{\rm n}$. At this temperature $R_{\rm s} = R_{\rm n} = R_{\rm n+s} = 1$. Again, as expected for $T > 1.8\,$K, when $\nu_{\rm n} < \nu'_{\rm s}$ (see Fig. 4b) we have $R_{\rm s} > 1$, $R_{\rm n} < 1$, meaning that the energy goes from the less damped normal component to the more damped superfluid one.

## 3 Turbulence of superfluid component in $^3$He

For future references we need the balance equation for the superfluid and normal fluid spectra $\mathcal{E}_{\rm s}(k,t)$ and $\mathcal{E}_{\rm n}(k,t)$ in the interval of scales, where the viscous dissipation is negligible. This equations directly follows from Eq. (4):

$$\frac{\partial \mathcal{E}_{\rm s}}{2\partial t} + \frac{\partial \varepsilon_{\rm s}}{\partial k} = \alpha \Omega(\mathcal{E}_{\rm ns} - \mathcal{E}_{\rm s})\,, \quad \frac{\partial \mathcal{E}_{\rm n}}{2\partial t} + \frac{\partial \varepsilon_{\rm n}}{\partial k} = \frac{\alpha \rho_{\rm s}}{\rho_{\rm n}} \Omega(\mathcal{E}_{\rm ns} - \mathcal{E}_{\rm n})\,. \tag{8a}$$

Here $\mathcal{E}_{\rm ns}(k,t)$ is the normal- and superfluid-velocity cross-correlation function.

Due to high kinematic viscosity of $^3$He (comparable to that of olive oil) $u_{\rm n} \ll u_{\rm s}$ and normal component can usually be considered at rest. In such circumstances, we approximate in our theoretical analysis the overall behavior of quantum turbulence as dominated by the superfluid velocity component only, Ref. [9,10]. If so, in steady state:

$$\frac{d\varepsilon_{\rm s}(k)}{dk} + \alpha\Omega\mathcal{E}_{\rm s}(k) = 0\,, \quad \varepsilon_{\rm s}(k) \simeq \frac{5}{8}k^{5/2}\mathcal{E}_{\rm s}^{3/2}(k)\ . \tag{8b}$$

Notice that rigorous derivation of energy flux over scale $\varepsilon_{\rm s}(k)$ involves the third-order correlation function of $\boldsymbol{u}_{\rm s}(k)$, exactly as in classical turbulence. To overcome this difficulty, we have employed in Eq. (8b) the simplest algebraic closure for $\varepsilon(k)$ via $\mathcal{E}_{\rm s}(k)$, suggested by Kovasznay [24]. The prefactor $\frac{5}{8}$ was chosen to get numerical value of the Kolmogorov constant reasonably close to its experimental value.

The ordinary differential Eq. (8b) was solved by L'vov-Nazarenko-Volovik (LNV) in Ref. [9] with the boundary conditions $\mathcal{E}_{\rm s}(k_0) = \mathcal{E}_0$ at the lowest wavenumber $k_0$ in the inertial interval:

$$\mathcal{E}_{\rm s}(k) = \mathcal{E}_0\Big(\frac{k_0}{k}\Big)^3\Big[\Big(1-\frac{\Omega}{\Omega_{\rm cr}}\Big)\Big(\frac{k}{k_0}\Big)^{2/3} + \frac{\Omega}{\Omega_{\rm cr}}\Big]^2, \quad \Omega_{\rm cr} = \frac{5}{4}\sqrt{k_0^3\,\mathcal{E}_0}\ . \tag{9}$$

At $\Omega = \Omega_{\rm cr}$ LNV-spectrum (9) becomes a scale-invariant "critical" spectrum $\mathcal{E}_{\rm cr}(k) = \mathcal{E}_0(k_0/k)^3$ .

For $\Omega < \Omega_{\rm cr}$, the "subcritical" (i.e. going *above* critical) solution of (9) consists of two parts. For $k$ smaller than the crossover wavenumber $k_\times = k_0\Big[\Omega/(\Omega_{\rm cr}-\Omega)\Big]^{3/2}$ it is close to the critical solution, while for $k \gg k_\times$, the subcritical spectrum (9) has K41 asymptote $\mathcal{E}_{\rm sb}(k) \Rightarrow \mathcal{E}_0\Big(\frac{k_0}{k}\Big)^{5/3}\Big(1-\frac{\Omega}{\Omega_{\rm cr}}\Big)^2$, with the energy flux $\varepsilon_\infty = \varepsilon_0(1-\Omega/\Omega_{\rm cr})^3$, smaller than the energy influx $\varepsilon_0$. The difference $\varepsilon_0 - \varepsilon_\infty$ is dissipated by the mutual friction.

For $\Omega > \Omega_{\rm cr}$, the "supercritical" (i.e. going *below* critical) spectrum (9) terminates at some final $k_* = k_0\Big[\Omega/(\Omega-\Omega_{\rm cr})\Big]^{3/2}$. However, if $\Omega$ slightly larger than critical, $\Omega-\Omega_{\rm cr} \ll \Omega_{\rm cr}$ and $k_* \gg k_0$, the supercritical spectrum for $k < k_*$ is close to the critical one.

One-fluid Sabra-shell model simulations of $^3$He turbulence [10], shown in Fig. 6a are in good agreement with the analytical results, Fig. 6b.

## 4 Thermally driven superfluid turbulence of $^4$He

In Sec. 2, devoted to mechanically driven turbulence of superfluid $^4$He, we showed that the corresponding energy spectra are very similar to spectra in classical turbulence, where main mechanism of the energy dissipation is molecular viscosity. We stressed that due to high correlation of normal and superfluid velocities, discussed in Sec. 2.3, the mutual friction plays secondary role leading

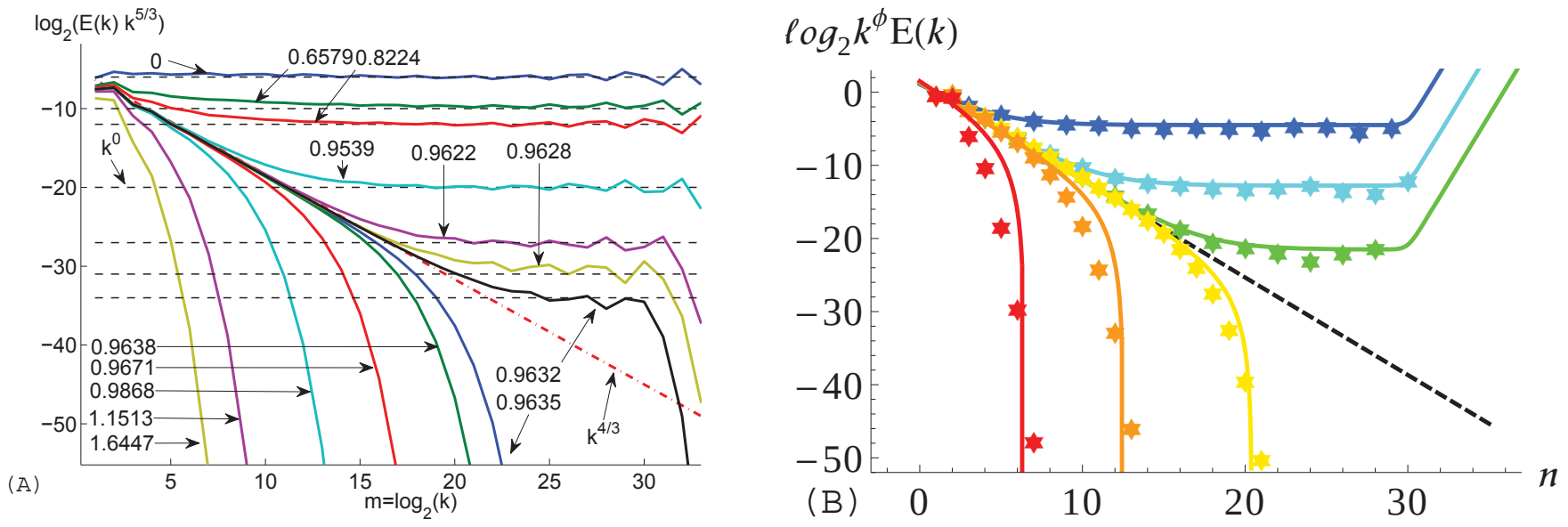

**Fig. 6** Panel a: Numerical results for the energy spectrum for different values of $\Omega/\Omega_{\rm cr}$. Panel b: Analytical functions $E_{\rm s}(k)$, Eq. (9) in the algebraic energy-flux model (8b). The stars show comparison with data from the Sabra-shell model at similar values of the ratio $\Omega/\Omega_{\rm cr}$. The bottleneck effect $E_{\rm sp}(k) \sim k^2$ at the highest wavevectors is removed from the numerical spectra by a proper choice of parameters to simplify the simulations.[10] (Color Online)

to subdominant effects in the high $k$-region, such as some energy redistribution between scales and intermittency enhancement.

The physics of superfluid turbulence in $^3$He with resting normal fluid, discussed in Sec. 3, is completely different. Here the main and much stronger mechanism of energy dissipation is due to mutual friction, resulting in the LNV-spectra (9) that decay much faster than the classical K41-spectrum (2b).

In this Sec. 4, we review results of Refs. [11, 25] and show that in thermally driven turbulence the relative role of energy dissipation due to molecular viscosity and due to mutual friction depends on the wavenumber $k$ and that resulting energy spectra share features of spectra in two previously discussed limiting cases. Namely, they are not of a classic K41 form (2b), but are partially suppressed by the mutual friction (to smaller extent than in $^3$He), leading to energy dissipation at all scales, enhanced by the counterflow-induced decoupling of the normal- and superfluid velocity fluctuations.

## 4.1 Experimental background

Our analytical model of thermal counterflow is inspired by two sets of experiments, discussed in following Sec. 4.1.1 and 4.1.2.

### *4.1.1 Prague triple-flow experiment*

Consider first reported in [11, 26] complementary experimental, numerical and theoretical study of turbulent coflow, counterflow and pure superflow of superfluid $^4$He in a channel, resulting in a physically transparent picture of decaying quantum turbulence that accounts for interactions of coexisting quantum and classical components of turbulent superfluid $^4$He.

For better comparison of physics of turbulence in these three types of flows in Ref. [11] it was used the same channel, sketched in Fig. 7a. The thermal counterflow was generated in a channel with one of its ends sealed and equipped with a heater and open at the other end to a superfluid helium bath (see Fig. 7a, left). Here both components move relative to the channel walls. The normal fluid carries the heat flux away from the heater, giving rise, due to conservation of mass, to a superfluid current in the opposite direction. In *pure superflow*, sketched in Fig. 7a, middle, superleaks (i.e., filters located at the channel end with sub-micron-sized holes permeable only to the inviscid superfluid component) allow a net flow of the superfluid component in the channel: $\boldsymbol{U}_{\rm s} \neq 0$, while the normal-fluid component is remaining, on average, at rest: $\boldsymbol{U}_{\rm n} = 0$. In both cases, the fields $\boldsymbol{u}_{\rm s}(\boldsymbol{r}, t)$ and $\boldsymbol{u}_{\rm n}(\boldsymbol{r}, t)$ are expected to be different at all scales. Thermal counterflow and pure superflow therefore represent two special cases of counterflow, characterized by non-zero difference in mean flow velocities of the superfluid and normal components. The classical-like mechanical forcing (e.g., by compressing a bellows) results in a coflow,

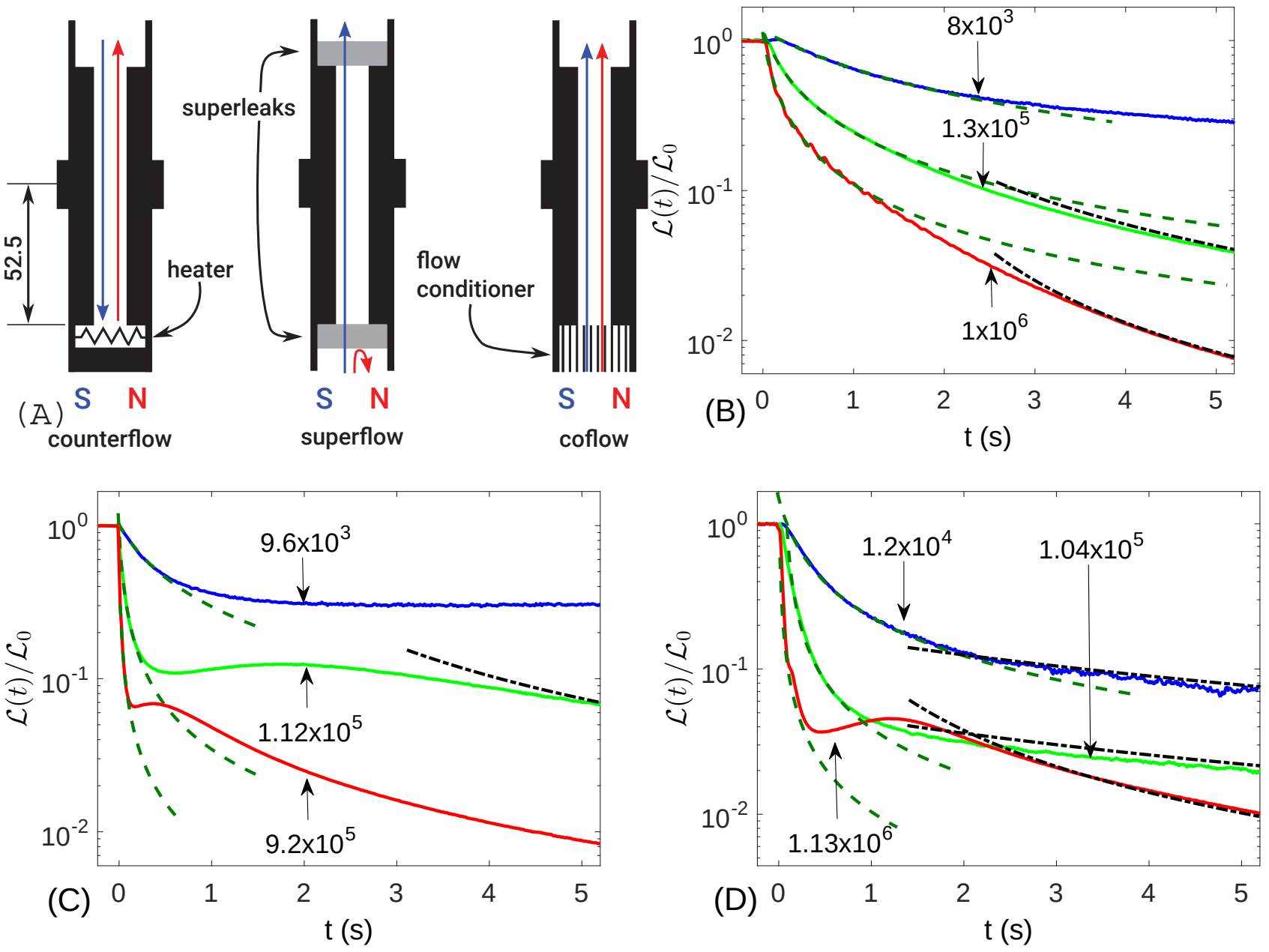

**Fig. 7** Panel a: Schematic presentation of experimental setup for the study of counterflow, pure superflow, and coflow. S and N stand for superfluid and normal components. Counterflow is produced thermally by a heater. Superflow and coflow are driven mechanically by bellows. The turbulence is probed in the middle of the channel by second-sound, excited and detected by mechanical vibration of a porous membrane. Panels b,c,d: Experimental data of VLD decay $\mathcal{L}(t)/\mathcal{L}_0$ in the coflow (b), counterflow (c) and superflow (d), normalized by initial VLD $\mathcal{L}_0$. The lines correspond (from top to bottom) to $\mathcal{L}_0 \simeq 10^4$ (blue lines), $10^5$ (green lines) and $10^6$ (red lines). The explicit values of $\mathcal{L}_0$ are shown in figures. Quantum $t^{-1}$-fits are shown by dashed dark green lines, while the classical $t^{-3/2}$-fits, – by black dot-dashed lines. [11] (Color Online)

the closest analogue to classical viscous channel flow, where both components move, on average, with the same velocity, see Fig. 7a, right.

In Figs. 7b,c,d we present typical experimental time dependencies of $\mathcal{L}(t)$, decaying by two-three orders of magnitude. The initial stage of decay in all three types of the flow, including the coflow regime, typically follows the form $\mathcal{L}(t) \propto 1/(t-\tau_1)$ that may be rationalized in the framework of the Vinen evolution equation for $\mathcal{L}(t)$ as decay of random tangle of quantized vortex lines. The energy spectrum of this vortex tangle $\mathcal{E}_s^Q(k)$ is dominated by the intervortex scales $\ell k \sim 1$. The late stage of the decay follows a $t^{-3/2}$. It is commonly believed that this dependence is caused by the classical Richardson-Kolmogorov cascade in the superfluid component, with $\mathcal{E}_s^{cl}(k) \propto k^{-5/3}$ spectrum.

A natural way to rationalize these observations is to assume that the resulting form of the turbulent energy spectrum of the superfluid component $\mathcal{E}_s(k)$ consists of two parts $\mathcal{E}_s(k) = \mathcal{E}_s^Q(k) + \mathcal{E}_s^{cl}(k)$: (i) the classical region spanning large scales from the integral length-scale $\Delta$ down to the intervortex distance $\ell$, where it is followed by (ii) a quantum contribution $\mathcal{E}_s^Q(k)$, corresponding to the random tangle of quantized vortex lines, having a form of a peak. Qualitatively, the energy spectrum in coflow is sketched in Fig. 8a, while the spectra in counterflow and superflow (Fig. 8b) have a form, qualitatively different from that in the coflow.

### *4.1.2 Tallahassee, Florida visualization experiment*

Crucial information on the steady-state statistics in counterflowing superfluid $^4$He was reported in Ref. [12]. The idea of the visualization experiment was to excite in the normal or superfluid $^4$He triplet-state of $^4$He$_2$ molecules by a focused femtosecond laser beam. This state has a radiative lifetime about 13 s. After a controlled time, the line of excited $^4$He$_2$ molecules can be imaged using laser-induced fluorescence. Data analysis [12, 13] allowed to measure streamwise projection of the normal fluid velocity $V_x(y)$ as a function of the wall normal direction $y$, see Fig. 9a and 9b. In [12, 13] these data was used to find second-order transversal velocity structure functions $S_2^\perp(y) = \left\langle [V_x(Y+y) - V_x(Y+y)]^2 \right\rangle_Y$. As seen in Fig. 9c, $S_2^\perp(y) \propto y$. Assuming isotropy of turbulence and using simple scaling arguments in the limit of very large scaling interval, the scaling $S_2^\perp(y) \propto y$ corresponds to the scaling of the 1D energy spectrum $\mathcal{E}(k) \propto 1/k^2$ [12]. The actual situation is not so simple. Nevertheless, this observation definitely shows important for us fact that in the counterflow turbulence $\mathcal{E}(k)$ decays faster with $k$ than in K41 prediction $\mathcal{E}(k) \propto k^{5/3}$.

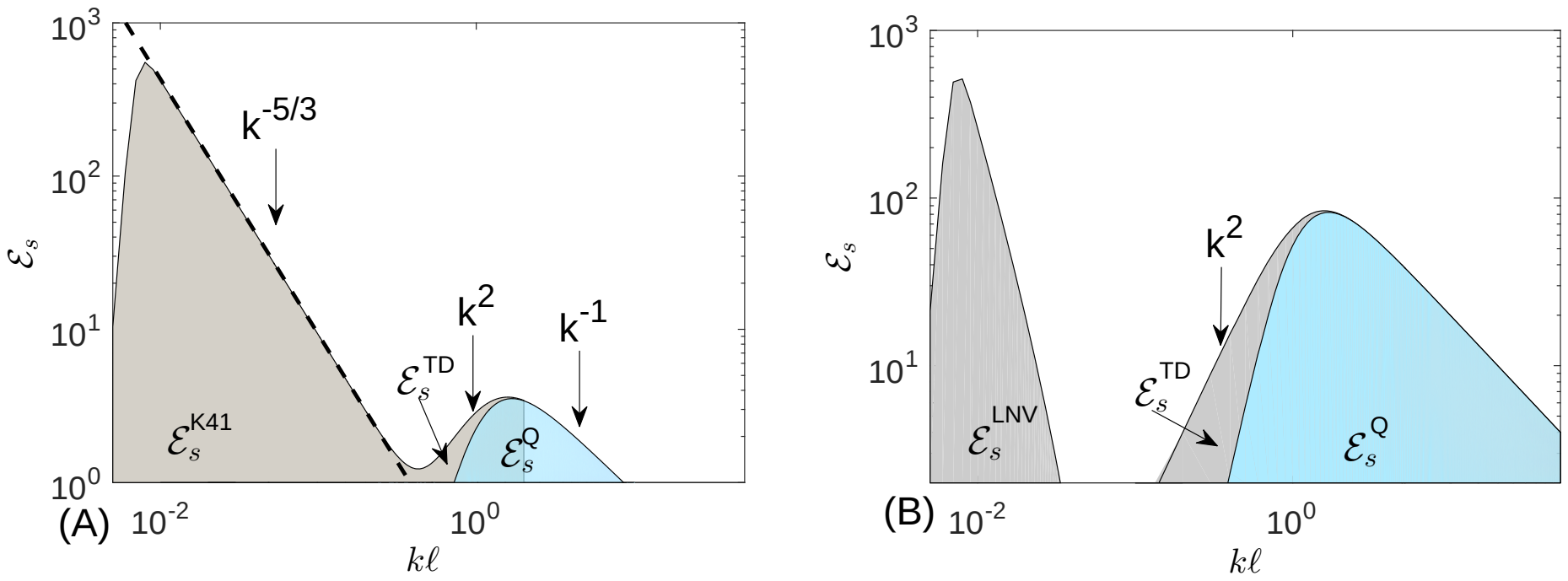

**Fig. 8** Sketch of the stationary superfluid turbulent energy spectrum in log-log coordinates, $\log \mathcal{E}_s(k)$ vs. $\log(k\ell)$. Spectrum $\mathcal{E}_s(k)$ consists of classical $\mathcal{E}_s^{cl}(k)$ and quantum $\mathcal{E}_s^Q(k)$ parts, colored in light blue and cyan. In coflow, (Panel a) $\mathcal{E}_s^{cl}(k)$ consists of cascade part $\mathcal{E}_s^{K41}(k) \propto k^{-5/3}$ (for $k < k_\times$) and thermodynamic equilibrium part $\mathcal{E}_s^{TD}(k) \propto k^2$ for $1/\ell \gtrsim k > k_\times$. In counterflow and pure superflow (Panel b), the quantum contribution $\mathcal{E}_s^Q(k)$ and the classical thermal bath part $\mathcal{E}_s^{TD}(k)$ look similar to that in coflow, while the cascade part, supercritical LNV-spectrum $\mathcal{E}_{LNV}$ Eqs. (9), ends at some $k_* < 1/\ell$ and does not provide energy to the quantum vortex tangle in stationary regime. [11] (Color Online)

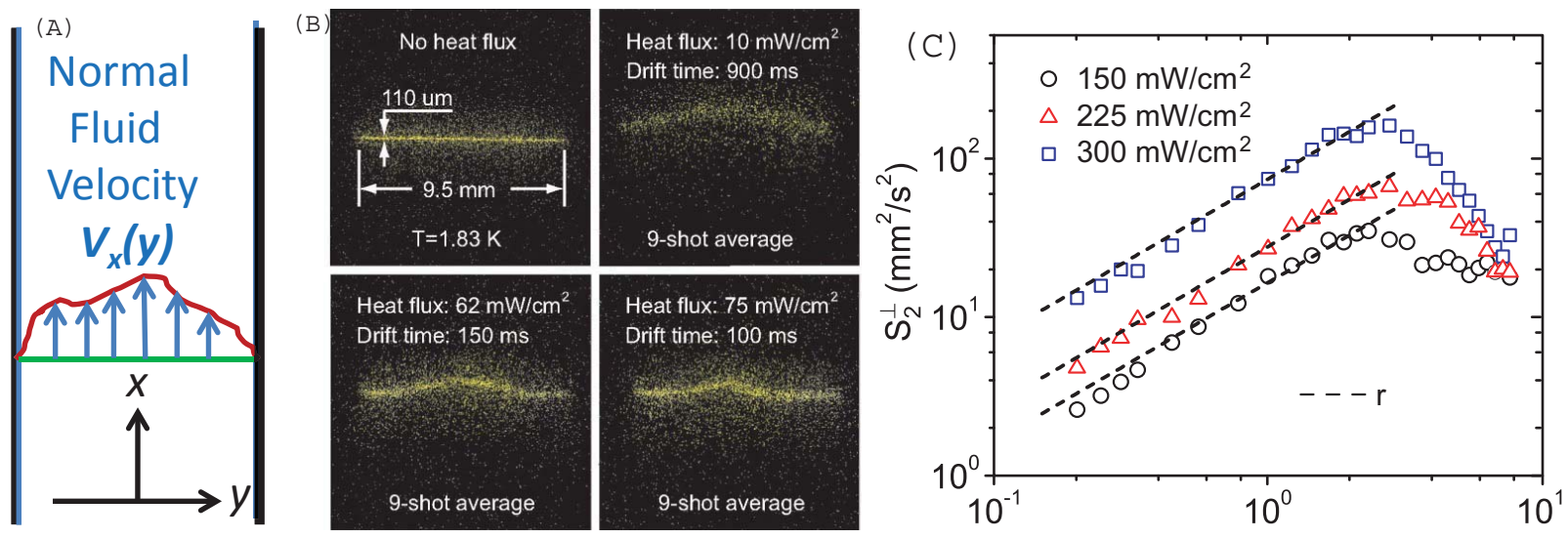

**Fig. 9** Panel a: Sketch of the channel in the counterflow experiment [12]. Panel b: Typical fluorescent images showing the motion $^4He_2$ tracer lines in thermal counterflow. Panel c: The observed in [12] steady-state second order structure function $S_2(r)$ of the normal-fluid turbulence with heat fluxes of 150, 225, and 300 mW/cm$^2$, respectively. The black dashed lines represent a power-law form of $S_2(r) \propto r$.(Color Online)

## 4.2 Toward a theory of quantum turbulence with counterflow

The first important ingredient of understanding of the steady-state energy spectrum in counterflow turbulence is decoupling of the normal and superfluid velocity fluctuations, $\boldsymbol{u}_\mathrm{n}(\boldsymbol{r},t)$ and $\boldsymbol{u}_\mathrm{s}(\boldsymbol{r},t)$, caused by the their sweeping in opposite direction by the corresponding mean velocities $\boldsymbol{U}_\mathrm{n}$ and $\boldsymbol{U}_\mathrm{s}$[25]. Considering the velocity fluctuations of characteristic scale $R$ (referred to here as $R$-eddies), one can easily estimate their overlapping time as $\tau_\mathrm{ov}(R) = R/U_\mathrm{ns}$.

Another characteristic time is the coupling time $\tau_\mathrm{cp}$, required to correlate the normal and superfluid motions in $R$-eddies by the mutual-friction. To estimate $\tau_\mathrm{cp}$, we use Eqs. (4) to see that right-hand-side of equation for $\partial(\boldsymbol{u}_\mathrm{s} - \boldsymbol{u}_\mathrm{s})/\partial t$ is proportional to $\approx \alpha\Omega(\boldsymbol{u}_\mathrm{s} - \boldsymbol{u}_\mathrm{s})$, where $\approx \tilde{\alpha} = \alpha\rho/\rho_\mathrm{n}$. This means that $\tau_\mathrm{cp} \simeq \pi / \approx \alpha\Omega$. If $\tau_\mathrm{ov} \ll \tau_\mathrm{cp}$, mutual friction does not have enough time to couple the normal and superfluid velocity fluctuations. In this case one expects that the velocities remain uncoupled. Otherwise, i.e. for $\tau_\mathrm{ov} \gg \tau_\mathrm{cp}$, full coupling of velocities is expected. Therefore, the coupling-decoupling process is governed by the dimensionless "decoupling" parameter $\zeta(k) = \tau_{\rm cp}/\tau_{\rm ov} \simeq kU_{\rm ns}/\Omega_{\rm ns}$ with $k \approx \pi/R$.

The analytical theory of the coupling-decoupling processes [25], using Langevin inspired approach to model the nonlinear term, results in the analytical expression for the cross-correlation $\mathcal{E}_{\rm ns}(k)$ :

$$\mathcal{E}_{\rm ns}(k) = D[\zeta(k)]\,\frac{\rho_{\rm s}\mathcal{E}_{\rm n}(k) + \rho_{\rm n}\mathcal{E}_{\rm s}(k)}{\rho}\,,\ D[x] = \frac{\arctan(x)}{x}\,,\ \zeta(k) = \frac{kU_{\rm ns}}{\approx \alpha\Omega}\,. \tag{10}$$

"Decoupling function" $D(x)$, is defined such that $D(x) = 1$ for $x = 0$ (i.e. for $k = 0$), $D(2) \simeq 0.5$ and it decays as $\pi/(2x)$ for large $x$.

Substituting $\mathcal{E}_{\rm ns}(k)$ into the balance Eqs. (8) and taking for simplicity $T \approx 1.95\,$K, for which $\rho_{\rm s} = \rho_{\rm n}$ and, as we expect $\mathcal{E}_{\rm n}(k) = \mathcal{E}_{\rm s}(k) \equiv \mathcal{E}(k)$, we get an ordinary differential equation for $\mathcal{E}(k)$

$$\frac{5}{8}\frac{d}{dk}\,k^{5/2}\mathcal{E}^{3/2}(k) = \Omega\,\mathcal{E}(k)\left[\frac{\arctan[\zeta(k)]}{\zeta(k)} - 1\right], \tag{11}$$

with the analytical solution [11]

$$\begin{aligned}
\mathcal{E}(k) &= \mathcal{E}_0\Big\{1 + A\left[I\Big(\frac{k}{k_\times^{\rm cf}}\Big) - I\Big(\frac{k_0}{k_\times^{\rm cf}}\Big)\right]\Big\}^2\Big(\frac{k_0}{k}\Big)^{5/3},\ A = \frac{8\,\Omega\,k_0^{2/3}}{15\,(k_\times^{\rm cf})^{2/3}\sqrt{k_0^3\mathcal{E}_0}}\,,\\
I(z) &= \frac{\sqrt{3}\pi}{5} + \frac{3}{20}\Big\{-\frac{4}{z^{2/3}} - 2\sqrt{3}\arctan[\sqrt{3} - 2z^{1/3}] + \frac{4\arctan[z]}{z^{5/3}}\\
&\quad -2\sqrt{3}\arctan[\sqrt{3} + 2z^{1/3}] + \ln\Big[1 - \frac{3z^{2/3}}{(1+z^{2/3})^2}\Big]\Big\},\quad k_\times^{\rm cf} = \frac{2 \approx \alpha\Omega}{U_{\rm ns}}\,.
\end{aligned} \tag{12}$$

For $k_0 < k \ll k_\times^{\rm cf}$ this spectrum deviates down from the K41 $\frac{5}{3}$-spectrum:

$$\mathcal{E}(k) \simeq \mathcal{E}_0\Big\{1 + \frac{A}{4\,(k_\times^{\rm cf})^{4/3}}\left[k_0^{4/3} - k^{4/3}\right]\Big\}^2\Big(\frac{k_0}{k}\Big)^{5/3}\,. \tag{13a}$$

For $k \gg k_\times^{\rm cf}$ it crucially depends on the value of $A$. There exists a critical value $A_{\rm cr} = 1/I(\infty) = 5/(\pi\sqrt{3}) \approx 0.92$. For $A < A_{\rm cr}$ the system asymptotically tends to K41 spectrum $\mathcal{E}(k) \simeq \mathcal{E}_0\Big\{1 - \frac{A}{A_{\rm cr}}\Big\}^2\Big(\frac{k_0}{k}\Big)^{5/3}$, but with the energy flux $\varepsilon_\infty^{\rm cf} = \varepsilon_0(1 - A/A_{\rm cr})^3$, smaller than the energy input rate $\varepsilon_0$. The difference $(\varepsilon_0 - \varepsilon_\infty^{\rm cf})$ is dissipated by mutual friction. This is similar to the subcritical LNV spectrum (9) of $^3$He turbulence with resting normal fluid component.

For $A > A_{\rm cr}$, $\mathcal{E}(k) = 0$ for large $k$. In the differential approximation used here, the spectrum $E(k)$ sharply terminates at some finite $k_*^{\rm cf}$, in the same manner as the supercritical LNV $^3$He spectrum:

$$\mathcal{E}(k) \propto \frac{1}{k^{5/3}}\Big\{\frac{1}{k^{2/3}} - \frac{1}{(k_*^{\rm cf})^{2/3}}\Big\}^2\,. \tag{13b}$$

The cutoff wave number $k_*^{\rm cf}$ may be found from the equation $1 = A[I(\zeta_{k_*^{\rm cf}}) - I(\zeta_0)] \approx A\,I(\zeta_{k_*^{\rm cf}})$. When $A \to A_{\rm cr}$, $k_*^{\rm cf} \to \infty$ and $\mathcal{E}(k) \propto k^{-3}$ at large $k$, exactly as in the critical LNV $^3$He spectrum.

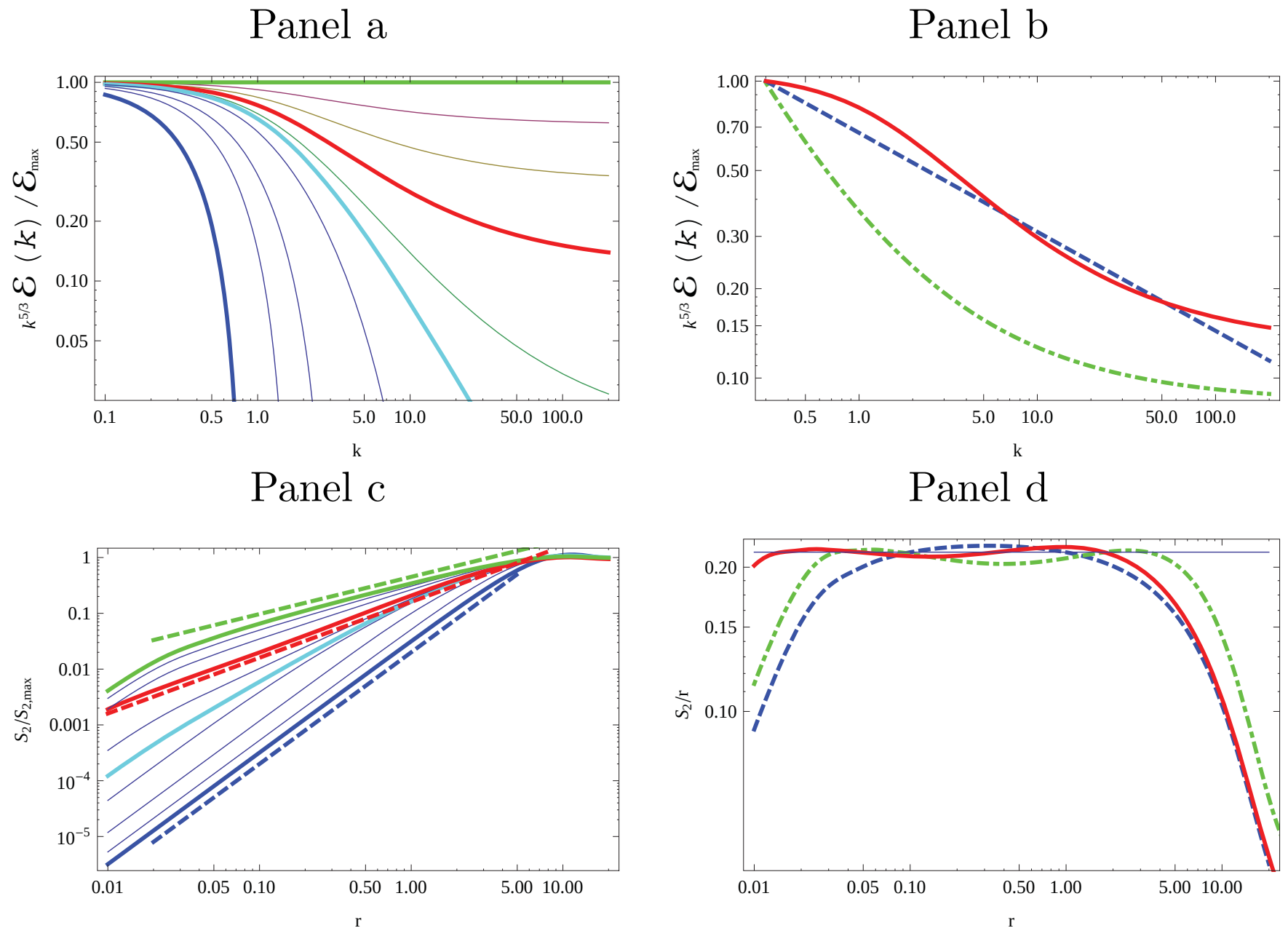

**Fig. 10** Panel (a) Log-Log plots of compensated normalized energy spectra $k^{5/3}\mathcal{E}(k)/\mathcal{E}_{\max}$ for $k_0 = 0$ and different values of $A$, starting with $A = 0$ (upper green line), through $A = 0.2,\ 0.4,\ 0.6$ (red thick line) $0.8, 0.92$ (thick cyan line – critical value), $1,\ 1.5,\ 2.0\,3.0$ and last line – $A = 6$. Panel (b) Log-Log plots of compensated normalized energy spectrum $k^{5/3}\mathcal{E}(k)/\mathcal{E}_{\max}$ for $k_0 = 0.3k_\times^{\rm cf}$ with $A = 0.6$ (solid red line), $1/k^2$ (after compensation by $k^{5/3}$) – blue dashed line and LNV subcritical spectrum $k^{-3}[1 + (k/k_*^{\rm cf})^{2/3}]^2$ with $k_*^{\rm cf} = 1.15k_\times^{\rm cf}$ – green dash-dotted line. Panel (c) Log-log plots of the second order structure functions $S_2(r)$ (normalized by their large $r$ limit), computed with Eq. (14) and with the same values of $A$ [and the same color code] as in Panel (a). The thick dashed straight lines indicate scaling laws: upper green line $\propto r^{2/3}$, middle red line $\propto r$ and lower blue line $\propto r^2$. Panel(d) The normalized compensated by $1/r$ structure functions $S_2(r)$, calculated using the spectra shown in Panel(b) (with the same color code). (Color online)

The energy spectra (12) for different values of $A$ in the range $0 \leqslant A \leqslant 6$, including the critical value $A_{\rm cr} \simeq 0.92$, are shown in Fig. 10a. We see that as mutual friction force, characterized by the dimensionless parameter $A$, becomes stronger, the energy spectra are progressively suppressed, going from K41 behavior $\mathcal{E}(k) \propto k^{-5/3}$ (for $A = 0$) towards the critical spectrum $\mathcal{E}(k) \propto k^{-3}$ at $A = A_{\rm cr} \simeq 1$. Further increase in $A$ localizes energy spectra in the $k$-space.

## 4.3 Counterflow structure functions and energy spectra

As we already discussed in Sec. 2.1, statistics of turbulence may be characterized in terms of velocity structure functions $S_n(R)$, Eq. (2a). In isotropic turbulence $S_2(R)$ can be expressed in terms of the energy spectra $\mathcal{E}(k)$:

$$S_2(r) \simeq \int_{k_0}^{k_{\max}} \mathcal{E}(k)[1 - \sin(kr)/(kr)]\, dk\,, \tag{14}$$

where for quantum turbulence in the superfluid component $k_{\max} \simeq \pi/\ell$. Taking for concreteness $k_0 \simeq 0.3k_\times^{\rm cf}$ and $k_{\max} \simeq 200\, k_\times^{\rm cf}$ [11], relevant for the Florida visualization experiment [12] we compute $S_2(r)$ with the set of spectra, shown inFig. 10a and plot the result in Fig. 10c. With the K41 spectrum (upper green solid line) we observe expected scaling behavior $S_2(r) \propto r^{2/3}$ in the interval of about two decades.

For nonzero values of $A$, the log-log plots of $S_2(r)$ vs $r$ can be considered as approximately straight lines with the slope that increases with $A$. In particular, for $A = 0.6$ $S_2(r)$, shown in Fig. 10c by solid red line, it is practically indistinguishable from the straight line with the slope +1 (shown by dashed red line) in the interval $0.02 < r < 2$. This means that, for $A = 0.6$, $S_2(r) \propto r$ with high accuracy in the interval of two decades. To see this better, we present in Fig. 10d the plot of $S_2(r)$ compensated by $1/r$. The solid red line in Fig. 10d indeed is very close to the black thin horizontal line.

Notice, that the energy spectrum for $A = 0.6$, used to find $S_2(r)$ (red solid line in Fig. 10b), is essentially different from the scale-invariant spectrum $\mathcal{E}(k) \propto k^{-2}$ (dashed blue line) that results in $S_2(r)$, shown in Fig. 10d by the dashed blue line. Unexpectedly, this result demonstrates scale-invariant behavior $S_2(r) \propto r$ on a shorter range. We also computed $S_2(r)$, using supercritical LNV spectrum (13b) with $k_*^{\rm cf} = 1.15k_0$. This spectrum, shown in Fig. 10b by the green dash-dotted line, is very different from the $1/k^2$ behavior (blue dashed line). Nevertheless, the resulting structure function $S_2(r)$ (green dash-dotted line) again demonstrates the scale-invariant behavior of $S_2(r) \propto r$ over more than two decades.

We conclude that very different energy spectra, including the spectrum (12) with $A \simeq 0.6$, found here, can result in the reported Ref. [12] $S_2(r) \propto r$ behavior with somewhat smaller extent of about one decade. This means that our analytical model does not contradict the scaling behavior of $S_2(r)$ [12]. It is worth to mention that the very same visualization data on streamwise velocity $u_x(y)$ dependence on the wall normal direction $y$ (used in [12] to compute $S_2(r)$) can be used to find directly $E(k_y) = \big\langle | \approx u_x(k_y)|^2 \big\rangle$ as suggested in the ongoing project [14].

From theoretical viewpoint, we should state that in spite of the similarity of these experimental observation and our analytical modeling, at this stage we are not in the position to claim that the model explains the observed simple behavior $S_2(r)$ and $E(k)$. The reason is that we adopted in our approach some uncontrolled approximations and simplifications, widely used and reasonably well justified in the studies of classical hydrodynamic turbulence, like isotropy and locality of the energy transfer over scales. Unfortunately, very little is know about validity of these approximations in case of thermally driven quantum counterflow turbulence.

## Summary

In this overview we tried to rise more questions than to give answers in order to demonstrate that much more experimental, analytical and numerical studies are required to achieve desired level of understanding and description of basic physical mechanisms, that governs quantum turbulence in superfluid He. In particular:

- In the mechanically driven $^4$He flow, the issues of intermittency, velocity coupling and energy exchange between the normal and superfluid subsystems call for further clarification;
- In the mechanically driven $^3$He flow, the high-order turbulent statistics and possible intermittency effects are still completely unknown;
- In $^4$He counterflow turbulence, the uncontrolled approximations used in presented analytical model of the energy spectra, have to be either better justified or relaxed.

**Acknowledgements** We acknowledge L. Skrbek, E. Varga, W. Guo and J. Gao who provided us with their experimental results prior to publicationsand allowed to use them in preparing this review. Useful discussion with them and with J. Vinen, I. Procaccia, S. Nazarenko, G. Volovik, C. Barenghi, P-E Roche and other colleagues made this review possible.